\RequirePackage{fix-cm}
\documentclass[smallextended]{svjour3}       
\usepackage[italian,english]{babel}
\smartqed  
\usepackage{amsmath}
\usepackage{amssymb}
\usepackage{lipsum}
\usepackage{dsfont}
\usepackage{multirow}
\usepackage{afterpage}
\usepackage{amsfonts}
\usepackage{bm}%
\usepackage{babel}
\usepackage{hyperref}
\usepackage{graphicx}
\usepackage{color}
\usepackage{graphicx}
\begin{document}

\title{Characterizing nonclassical correlation using affinity 
}


\author{R. Muthuganesan        \and
       V. K. Chandrasekar 
}


\institute{Centre for Nonlinear Science  \& Engineering, School of Electrical \& Electronics Engineering, SASTRA Deemed University, Thanjavur - 613 401, Tamil Nadu, India\at
              Tel.: +91-8760372825\\
              \email{rajendramuthu@gmail.com}           
           \and
          Centre for Nonlinear Science  \& Engineering, School of Electrical \& Electronics Engineering, SASTRA Deemed University, Thanjavur - 613 401, Tamil Nadu, India\at
              \email{chandru25nld@gmail.com}
}

\date{Received: date / Accepted: date}

\maketitle

\begin{abstract}
Geometric discord (GD), a measure of quantumness of bipartite system, captures minimal nonlocal effects of a quantum state due to locally invariant von Neumann projective measurements. The original version of GD is suffered by local ancilla  problem. In this article, we propose a new version of geometric discord using affinity. This quantity satisfies all the necessary criteria of a good measure of quantum correlation for bipartite system and resolves local ancilla problem. We evaluate analytically the proposed discord for both pure and mixed states.  For an arbitrary pure state, it is shown that affinity--based geometric discord is same as geometric measure of entanglement. Further, a lower bound of this measure for $m \times n$ dimensional arbitrary mixed state and  a closed formula of proposed version of geometric discord for $2 \times n$ dimensional mixed state are obtained. Finally, as are illustrations, we have studied the quantum correlation of Bell diagonal state, isotropic  and Werner states.

\keywords{Nonlocality \and Affinity \and Projective measurements \and Quantum correlation}
\end{abstract}

\section{Introduction}
\label{intro}

Quantum correlations are peculiar nonlocal features of quantum mechanics which make a fundamental departure from the classical realm. They also regarded as valuable resources needed in quantum algorithms \cite{Datta2005,Datta2008}and communication protocols \cite{Cavalcanti,Madhok} which reveal the quantum advantages over their classical counterpart\cite{Nielsen2010}. Entanglement is believed to be one of the manifestations of nonlocality  of the quantum state \cite{vedral,Brunner}. Entanglement has been studied for various physical systems \cite{RLF2016,ASRAB,Franco} and its dynamics under various environments \cite{Xu2013,Mortezapour,Mortezapour2,Dijkstra,Arrigo,Orieux,Man2015}. To probe information about nonlocal content of a system, whether classical or quantum, one has to perform measurement. A notable key difference between classical and quantum behavior is the characteristics of measurements: while classical measurement can extract information without disturbing in principle the system under consideration, quantum measurement often unavoidably affects the measured system. Recently, researchers paid wide attention on correlation measures based on the quantum measurements some of the measures  introduced are  quantum discord \cite{Ollivier2001,Modi}, measurement-induced disturbance (MID) \cite{Luo2008}, geometric discord (GD) \cite{Luo2010,Bromley}, measurement-induced nonlocality (MIN) \cite{SLUO2011,Muthu1,Hu2015} and uncertainty-induced nonlocality (UIN) \cite{UIN}. Dynamics of quantum correlation (beyond entanglement) is also studied under various circumstances \cite{Bellomo2011,Muthu2,Muthu3,Aaronson,Cianciaruso,Haikka}. 

On the other hand, distance measures which quantify the closeness or similarity of two states in the state space play a central role for the classification of states in information theory. They are also associated with measure of geometric version of quantum correlations. Further, they are useful to quantify how precisely a quantum channel can transmit information. In this context, distance measures like trace distance, Hilbert-Schmidt norm, Jensen-Shannon divergence, fidelity induced metric and Hellinger distance are introduced \cite{Bengtsson,Jozsa2015}. Recently, measurement based nonclassical correlations (a family of discord-like measures) have been characterized using various distance measures \cite{Spehner}.  In the same spirit, quantum correlations using affinity-based metric is  a figure of central merit in the quantum estimation theory. The amount of quantum correlation in terms of affinity can be used as a resource for information processing. 

In this paper, we propose affinity--based quantum correlation measure (geometric discord) for bipartite system. It is shown that this quantity is a remedy to the local ancilla problem of Hilbert-Schmidt norm based geometric discord \cite{Luo2010}. We compute the proposed measure analytically for both arbitrary pure and $2 \times n $ mixed states. For pure state, we show that the affinity--based discord equals to Hilbert--Schmidt distance discord, modified geometric discord and geometric measure of entanglement. Further, we present a lower bound for arbitrary $m \times n$ dimensional mixed state in terms of eigenvalues of correlation matrix. To check the validity of our proposed quantity, we study the geometric discord for well--known families of two-qubit mixed states and compare with the original measure.

This paper is organized  as follows: In sec. \ref{correlation}, we briefly review the concept of geometric discord and local ancilla problem. In sec. \ref{affinity}, to cope with the issue of Hilbert--Schmidt norm discord, we introduce affinity--based discord and evaluate analytically  for both pure and mixed states. In sec. \ref{Example}, we study the geometric discord of well-known two qubit states. Finally, the conclusions are presented in sec. \ref{Concl}. 

\section{Geometric Discord}
\label{correlation}
Quantum discord, a first measure of  quantum correlations of bipartite system beyond entanglement, is initially introduced by Ollivier  and Zurek \cite{Ollivier2001}. It is defined as minimal loss of information due to von Neumann projective measurements on a given state $\rho$ i.e., 
\begin{align}
  Q(\rho)=~^{\text{min}}_{\{ \Pi_k^{A}\} } \{ I(\rho)-I[\Pi^{A}(\rho)]\}.
\end{align}
where the minimum is taken over the von Neumann projective measurements $\{ \Pi_k^{A}\}$ on party $A$. Here $I(\rho):=S(\rho^A)+S(\rho^B)-S(\rho)$ is the quantum mutual information, $S(\rho):=-\text{Tr}(\rho \text{ln}\rho)$ is the von Neumann entropy of $\rho$, $\rho^A$ and $\rho^B$ are the marginal states of corresponding parties $A$ and $B$ respectively, $\text{Tr}(\cdot)$ denotes trace of a matrix and
\begin{align}
 \Pi^{A}(\rho)= \sum_k(\Pi_k^{A}\otimes \mathds{1})\rho (\Pi_k^{A}\otimes \mathds{1}) \label{postmeasurement}
\end{align}
is the resultant state after the projective measurements $\Pi_k^{A}=|k\rangle \langle k|$ on state space $A$ and $\mathds{1}$ is the identity operator on state space $B$. The operational meaning of quantum discord is nicely interpreted as the minimal decrement of entropy (information) due to the measurements. The presence of discord in separable state shows quantum advantageous even in the absence of entanglement.  It is shown that quantum discord is not calculable in closed form for an arbitrary two-qubit state \cite{Girolami2011} and is an NP-problem \cite{Huang}. However, recently the closed formula of quantum discord for higher dimensional X-state is investigated \cite{Rau}. 

The geometric version of discord is introduced using Hilbert-Schmidt norm as \cite{Dakic2010}
\begin{align}
  Q_G(\rho):=~^{\text{min}}_\chi ~\lVert \rho-\chi\rVert^2,
\end{align}
where the minimum is taken over set of all zero-discord states i.e., $Q_G(\chi)=0$. Further, Luo and Fu have shown the equivalence between the zero-discord and post-measurement state (\ref{postmeasurement}), and the definition of geometric discord (GD) is reformulated as \cite{Luo2010}
\begin{align}
D_G(\rho)=~~^{\text{min}}_{\{ \Pi_k^{A}\}} \lVert \rho-\Pi^{A}(\rho)\rVert^2.  \label{GD2}
\end{align}
Here also the optimization is taken over von Neumann projective measurements.  The merits of this quantity are easy to compute and experimentally realizable \cite{Jin,Passante,GirolamiPRL}. However, this quantity is having unwanted property of a quantum correlation measure, which is indicated by M. Piani \cite{Piani2012}. The problem of this quantity is that it may change rather arbitrarily through some trivial and uncorrelated action on the unmeasured party $B$. 

Consider a simple channel $\tau ^\sigma: X\rightarrow X\otimes\sigma$, i.e., the channel introduces a noisy ancillary state $\sigma$. Under such an operation 
\begin{align}
\| X\| \rightarrow \| \tau  ^\sigma X \|=\| X\| \sqrt{\text{Tr}\sigma^2},
\end{align}
since the Hilbert--Schmidt norm is multiplicative on tensor products. If $\rho$ is a bipartite state of our interest and $\rho^C$ is an ancilla state due to some channel, the resultant state may be denoted as $\rho^{A:BC}$. Then the GD of resultant state is
\begin{align}
D_G(\rho^{A:BC}) = D_G(\rho^{AB}) \text{Tr}(\rho ^{C})^2  \nonumber
\end{align}
implying that geometric discord differs arbitrarily due to local ancilla  as long as $\rho^{C}$ is mixed. Thus, adding or removing local ancilla -- a local and reversible operation -- adds or removes a factor of purity of the ancillary state to the GD of resultant state. This is known as local ancilla problem.

Luo and Fu remedied this local ancilla problem by replacing density matrix by its square root i.e., mathematically the remedied discord is defined as \cite{Chang2013}
\begin{equation}
 D_G^R(\rho ) =~^{\text{min}}_{\Pi ^{A}}\| \sqrt{\rho} - \Pi ^{A}(\sqrt {\rho} )\| ^{2},
\end{equation}
where $\rho^{AB}$ is recognized as $\rho$. Moreover, this problem can also be circumvented by modifying the definition of GD with choice of various distance measures. In what follows, we define a new variant of discord to resolve this issue and its properties in detail. 
\section{Affinity based geometric discord}
\label{affinity}
Affinity, like fidelity characterizes the  closeness of two quantum states.  Let $\mathcal{H}$ be an $n-$ dimensional Hilbert space and $\mathcal{L}(\mathcal{H})$ be the set of density matrices. For any $\rho$, $\sigma$ $\in \mathcal{L}(\mathcal{H})$, affinity is defined as \cite{Luo2004},  

\begin{align}
  \mathcal{A}(\rho,\sigma)=\text{Tr}(\sqrt{\rho}\sqrt{\sigma}).
\end{align}
This definition is much similar to Bhattacharya coefficient between two classical probability distribution \cite{Bhattacharyya}. It is worthful to mention that affinity is possess all the properties of Uhlmann--Jozsa fidelity \cite{Jozsa2015}. This quantity is more useful in quantum detection and estimation theory. Though affinity itself is not a metric. However, due to monotonicity and concavity property of affinity \cite{Luo2004}, one can define  any monotonically decreasing function of affinity  as a metric in state space. One such  affinity-based metric is defined as 
\begin{align}
  d_{\mathcal{A}}(\rho,\sigma)=\sqrt{1-\mathcal{A}(\rho,\sigma)}.
\end{align}
Based on the above metric, we introduce a new version of geometric measure quantum correlation (geometric discord). Defining geometric discord in terms of affinity as 
\begin{align}
D(\rho)=~^\text{{min}}_{\{ \Pi_k^{A}\} }\sum_k D_{\mathcal{A}}[\rho, (\Pi_k^{A}\otimes \mathds{1})\rho (\Pi_k^{A}\otimes \mathds{1})], \label{GD}
\end{align}
where $D_{\mathcal{A}}[\rho, (\Pi_k^{A}\otimes \mathds{1})\rho (\Pi_k^{A}\otimes \mathds{1})]= d^2_{\mathcal{A}}(\rho,\Pi^{A}_k(\rho))$. In principle, one can generalize this definition for multipartite scenario. Recently, the quantum discord of Gaussian state is studied using Hellinger distance interms of affinity \cite{Marian}, and this quantity is just twice of the measure in Eq.(\ref{GD}). 

Here, we demonstrate some interesting properties of affinity based geometric discord $D(\rho)$: 
\begin{enumerate}
\item[(i)]  $D(\rho)$ is non-negative i.e., $D(\rho)\geq 0$. 

\item[(ii)] $D(\rho)=0$ for any product state $\rho=\rho_{A}\otimes  \rho _{B}$ and the classical-quantum state in the form $\rho =\sum _{k}p_{k}|k\rangle \langle k| \otimes   \rho_{k}  $ with nondegenerate marginal state $\rho^{A}=\sum_{k}p_{k}|k\rangle \langle k|$.
For any product and classical-quantum state, one can always find $\Pi_k^{A}$ such that $\rho=\Pi^{A}(\rho)$  and $\mathcal{A}(\rho,\Pi^{A}(\rho))=1$, which leads zero discord.

\item[(iii)] $D(\rho)$ is locally unitary  invariant i.e., $D\left((U\otimes   V)\rho  (U\otimes   V)^\dagger\right)=D(\rho)$ for any local unitary operators $U$ and $V$. We have 
\begin{align}
D\left((U\otimes V)\rho  (U\otimes V)^\dagger\right)=&~^\text{{min}}_{\{ \Pi_k^{A}\} }\sum_k D_{\mathcal{A}}\big[ \left((U\otimes V)\rho  (U\otimes V)^\dagger\right),\notag \\
&(\Pi_k^{A}\otimes \mathds{1}) \left((U\otimes V)\rho  (U\otimes V)^\dagger\right) (\Pi_k^{A}\otimes \mathds{1})\big]  \nonumber \\
=& ~^\text{{min}}_{\{ \Pi_k^{A}\} }\sum_k D_A\big[\rho, (U\Pi_k^{A}U^\dagger\otimes\mathds{1})\rho(U\Pi_k^{A}U^\dagger\otimes\mathds{1})\big]  \nonumber\\
=& D(\rho).    \nonumber
\end{align}
\item[(iv)] For any $m \times n$ dimensional pure maximally entangled state with $m\leq n$, $D(\rho) $ has the maximal value of $\frac{m-1}{m}$. 
\item[(v)] $D(\rho)$ is invariant under the addition of any local ancilla to the unmeasured party $B$. 
\end{enumerate}

\subsection{\bf Remedying local ancilla problem}

{\bf Theorem 1:} \textit{The affinity based quantum correlation measure is invariant under addition of local ancilla}.
 
Proof: To prove this we first recall multiplicative property of affinity. For any $\rho_i$, $\sigma_j \in \mathcal{L} (\mathcal{H})$ $(i,j=1,2)$
\begin{align}
\mathcal{A}(\rho_1 \otimes \rho_2,\sigma_1 \otimes \sigma_2)=\mathcal{A}(\rho_1 \otimes\sigma_1)\cdot \mathcal{A}(\rho_2 \otimes\sigma_2). \label{multiaffinity}
\end{align}
After the addition of local ancilla the affinity between the pre-- and post--measurement state is
\begin{align}
\mathcal{A}\left(\rho^{A:BC},\Pi ^{A}(\rho^{A:BC})\right) = \mathcal{A}\left(\rho^{AB}\otimes  \rho ^{C},\Pi ^{A}(\rho^{AB})\otimes  \rho ^{C}\right). \nonumber
\end{align}
Using multiplicativity property of affinity Eq.(\ref{multiaffinity}), we have 
\begin{align}
 \mathcal{A}\left(\rho^{A:BC},\Pi ^{A}(\rho^{A:BC})\right) =& \mathcal{A}\left(\rho^{AB}, \Pi ^{A} (\rho ^{AB})\right)\cdot \mathcal{A}(\rho^{C},\rho^{C}) \nonumber  
\end{align}
and the affinity of same state is unity, which completes the proof. Then the affinity--based geometric discord is regarded as a good measure of quantumness of the system. 
\subsection{\bf Pure State}
{\bf Theorem 2:} \textit{For any pure bipartite state with Schmidt decomposition $| \Psi \rangle =\sum_{i}\sqrt{s_{i}}| \alpha _{i} \rangle \otimes | \beta _{i}\rangle $,}
\begin{equation}
 D(| \Psi \rangle\langle \Psi| )=1- \sum_{k} s_{k}^{2}.
\end{equation}
Proof: Using the identity $\Pi^A f(\rho)\Pi^A=f(\Pi^A \rho \Pi^A)$ \cite{GirolamiPRL}, one can rewrite the definition of GD using affinity as 
\begin{align}
D(\rho)=~1-^\text{{max}}_{\{ \Pi_k^{A}\} } \sum_k\text{Tr}[ \sqrt{\rho} (\Pi_k^{A}\otimes \mathds{1})\sqrt{\rho}(\Pi_k^{A}\otimes \mathds{1})]. \label{identity}
\end{align}
Noting that 
\begin{equation}
\rho=\sqrt{\rho}= | \Psi \rangle \langle \Psi| = \sum_{ij}\sqrt{s_{i}s_{j}}| \alpha_{i} \rangle \langle \alpha_{j}| \otimes  | \beta_{i} \rangle \langle \beta_{j}|. \nonumber  
\end{equation}
The von Neumann projective measurements on the subsystem $A$ can be expressed as $\Pi ^{A}=\{\Pi ^{A}_{k}\otimes \mathds{1}\} = \{| \alpha_{k}\rangle \langle \alpha_{k}|\otimes| \mathds{1}\} $ do not alter the marginal state $\rho^{A}$. The marginal state $\rho^A=\Pi^A(\rho^A)=\sum_k\Pi_k^A\rho^A\Pi_k^A$ is written as 
\begin{equation}
 \rho^{A}=\sum_{k} U| \alpha_{k}\rangle \langle \alpha_{k}| U^{\dagger}\rho^{A} U| \alpha_{k}\rangle \langle \alpha_{k}| U^{\dagger}.  \nonumber
\end{equation}
This marginal state $\rho^{A}$ can be written as spectral decomposition in the orthonormal bases $\{U| \alpha_{k}\rangle\} $ as
 \begin{equation}
 \rho^{A}= \sum_{k} \langle \alpha_{k}| U^{\dagger}\rho^{A} U| \alpha_{k}\rangle U| \alpha_{k}\rangle \langle \alpha_{k}| U^{\dagger}.
\end{equation}
After a straight forward calculation and simplification, we show that 
\begin{align}
\sum_k\text{Tr}[ \sqrt{\rho} (\Pi_k^{A}\otimes \mathds{1})\sqrt{\rho}(\Pi_k^{A}\otimes \mathds{1})]=\sum_k(\langle \alpha_{k}| U^{\dagger}\rho^{A} U| \alpha_{k}\rangle)^2=\sum_k s_k^2,  \label{resultT1}
\end{align}
where $\langle \alpha_{k}| U^{\dagger}\rho^{A} U| \alpha_{k}\rangle=s_k$ are the eigenvalues of state $\rho^A$. Substituting Eq.(\ref{resultT1}) in Eq.(\ref{identity}), we obtain the affinity based geometric discord for pure state as,
\begin{equation}
 D(| \Psi \rangle\langle \Psi| )=1- \sum_{k} s_{k}^{2}
\end{equation}
and hence theorem is proved. It is worthful to mention that for pure state, the proposed quantity is equal to earlier quantities such as skew information, Hilbert-Schmidt norm based discord, geometric measure of entanglement and remedied version of GD. For pure $m \times n-$ dimensional entangled state with $m\leq n$, the quantity $\sum_{k}s^{2}_{k}$ is bounded by $1/m$ and thus
\begin{equation}
D\left(| \Psi \rangle \langle \Psi|\right)\leq \frac{m-1}{m}
\end{equation}
with equality holds for maximally entangled state.

\subsection{\bf Mixed state}
Let $\{X_{i}:i=0,1,2,\cdots,m^{2}-1\} \in \mathcal{L}(\mathcal{H}^A)$ be a set of orthonormal operators for the state space $\mathcal{H}^A$ with operator inner product $\langle X_{i}| X_{j}\rangle = \text{Tr}(X_{i}^{\dagger}X_{j})$. Similarly, one can define $\{Y_{j}:j=0,1,2,\cdots,n^{2}-1\}  \in \mathcal{L}(\mathcal{H}^B)$ for the state space $\mathcal{H}^B$. The operators $X_{i}$ and $Y_{j}$ are satisfying the conditions $\text{Tr}(X_{k}^{\dagger }X_{l})=\text{Tr}(Y_{k}^{\dagger}Y_{l})=\delta _{kl}$. With this, one can construct a set of orthonormal operators $\{X_{i} \otimes Y_{j} \}\in \mathcal{L} (\mathcal{H}^{A}\otimes \mathcal{H}^{B}) $ for the composite system. Consequently, an arbitrary $m\times n$ dimensional state of a bipartite composite system can be written as
\begin{equation}
\sqrt{\rho}= \sum_{i,j}\gamma _{ij}X_{i}\otimes  Y_{j}, \label{c}
\end{equation}
where $\Gamma=(\gamma _{ij}) =\text{Tr} (\sqrt{\rho} ~X_{i}\otimes  Y_{j})$ are real elements of correlation matrix. For any orthonormal basis  $\{| k\rangle :k=0,1,2,\cdots,m-1\} $, $| k\rangle \langle k| =\sum_{i}a_{ki}X_{i}$ with $a_{ki}=\text{Tr}(| k\rangle \langle k|X_{i})$.

{\bf Theorem 3:} \textit{For any arbitrary bipartite state Eq. (\ref{c}), affinity--based geometric discord has a tight lower bound as
\begin{align}
D(\rho)\geq 1-\sum_i\mu _i,
\end{align}
where $ \mu _i$ are the eigenvalues of matrix $\Gamma \Gamma^t$ arranged in decreasing order and $t$ denotes the transpose of the matrix}.

{\bf Proof}: To prove this theorem, we first compute the affinity between the  pre- and post- measurement state. For the given bipartite state, we have 
\begin{align}
\text{Tr}[ \sqrt{\rho} (\Pi_k^{A}\otimes \mathds{1})\sqrt{\rho}(\Pi_k^{A}\otimes \mathds{1})]=\sum_{ii',jj'}\gamma_{i'j'}\gamma_{ij} r_{ki}X_i|k\rangle \langle k|\otimes Y_jY_j'.
\end{align}
Thus
\begin{align}
\sum_{k=1}^m\text{Tr}[ \sqrt{\rho} (\Pi_k^{a}\otimes \mathds{1})\sqrt{\rho}(\Pi_k^{a}\otimes \mathds{1})]=\sum_{ii',jk}r_{ki}\gamma_{i'j}\gamma_{ij} r_{ki},
\end{align}
\begin{align}
=\text{Tr}(R\Gamma\Gamma^tR^t).
\end{align}
From Eq.(\ref{identity}), affinity based quantum correlation measure can be written as 
\begin{align}
D(\rho)=1-~^\text{max}_{~R}~\text{Tr}(R\Gamma\Gamma^tR^t).
\end{align} 
If the marginal state $\rho^A$ is nondegenerate, the optimization is not required. For degenerate state we try the optimized $D(\rho)$ for any arbitrary mixed state. Putting $X_0=\mathds{1}/\sqrt{m}$ with real coefficients $\{ r_{ki}:i=0,1,2,\cdots, m^2-1\} $ for the expanding operator in the orthonormal basis $\{ X_i: i=0,1,2,\cdots, m^2-1\} $ of state space $A$. Now we have
\begin{align}
\sum_{i=0}^{m^{2}-1}r_{ki}r_{k^{'}i}=\text{Tr}\left(| k\rangle \langle k| k^{'}\rangle \langle k^{'}| \right)=\delta _{kk^{'}} \nonumber
\end{align}
with $r_{k0}=1/\sqrt{m}$. For $k=k^{'}$
\begin{align}
\sum_{i=1}^{m^{2}-1}r_{ki}^{2}= \frac{m-1}{m} \label{equalk} 
\end{align}
and for $k\neq k^{'}$
\begin{align}
\sum_{i=1}^{m^{2}-1}r_{ki}r_{k^{'}i}= -\frac{1}{m}.  \label{notequalk}
\end{align}
From the above Eqs. (\ref{equalk}) and (\ref{notequalk}) we can observe that $RR^t$ is a real matrix with eigenvalues $0$ and $1$ (of multiplicity $m-1$). By some mathematical manipulation, we obtain
the optimization as
\begin{align}
~^\text{max}_{~R}~\text{Tr}(R\Gamma\Gamma^tR^t)=~^\text{~~~max}_{C:CC^t=\mathds{1}} \text{Tr}(C\Gamma\Gamma^tC^t)=\sum_{i=0}^{m-1} \mu _i,
\end{align}
which leads the lower bound for affinity based discord and completes the proof of theorem 3. 

Next, we compute the closed formula for any $2 \times n$ dimensional system. The orthonormal  basis of the system $A$ is $\{ X_i\} =\{ \mathds{1}/\sqrt{2}, \sigma_1/\sqrt{2},\sigma_2/\sqrt{2},\sigma_3/\sqrt{2}\} $ and the marginal state can be written as 
\begin{align}
\rho^A=\frac{1}{\sqrt{2}} \frac{\mathds{1}^A}{\sqrt{2}}+\sum^3_{i=1}\text{r}_{ki} \frac{\sigma_i}{\sqrt{2}} \nonumber
\end{align}
with $\mathbf{r}=(r_{ki}=\text{Tr}(|k\rangle \langle k| \sigma_i))$ with $\lVert \mathbf{r}\rVert=1$. Now the matrix 
\begin{align}
R=\frac{1}{\sqrt{2}}
\begin{pmatrix}
1 & \mathbf{r}\\
1 & -\mathbf{r}
\end{pmatrix} \nonumber
\end{align} 
and partition $\Gamma$ as 
\begin{align}
\Gamma=
\begin{pmatrix}
\mathbf{v} \\
Z
\end{pmatrix}, \nonumber
\end{align} 
where $Z=(\gamma_{ij})_{i=1,2,3; j=0,1,2, \cdots, n^2-1}$ is a $3 \times n^2$ dimensional matrix, $\mathbf{v}=(\gamma_{00},\gamma_{01}, \gamma_{02}, \cdots,\gamma_{0n^2-1})$ is a row vector. The direct multiplication gives 
\begin{align}
\text{Tr}(R\Gamma\Gamma^tR^t)=\lVert \mathbf{v}\rVert ^2+\mathbf{r}ZZ^t\mathbf{r}^t. \nonumber
\end{align}
Then,the closed formula of discord is 
\begin{align}
D(\rho)=1-\lVert \mathbf{v}\rVert ^2-z_{\text{max}}, \label{Aresult}
\end{align}
where $z_{\text{max}}$ is maximal eigenvalue of matrix $ZZ^t$.
\section{Examples}
\label{Example}
In order to show the consistency in capturing quantum correlation of proposed quantity, we compute the affinity-based GD for well-known two qubit Bell diagonal state and Werner state and compare with original version of geometric discord.

\textit{Bell diagonal state}: The Bloch representation of the state can be expressed as 
\begin{equation}
\rho^{BD}=\frac{1}{4}\left[\mathds{1}\otimes\mathds{1}+\sum^3_{i=1}c_i(\sigma^i \otimes \sigma^i)\right]=\sum_{a,b}\lambda_{ab} |\beta_{ab}\rangle \langle \beta_{ab}|,
\end{equation}
where the vector $\vec{c}=(c_1,c_2,c_3) $ is a three dimensional vector composed of correlation coefficients such that $-1\leq c_i=\langle \sigma^i \otimes\sigma^i \rangle \leq 1$ completely specify the quantum state and $\lambda_{a,b}$,  here $a, b\in \{ 0,1\}$ denotes the eigenvalues of Bell diagonal state which are given by
\begin{equation}
\lambda_{a,b}=\frac{1}{4}\left[1+(-1)^a c_1-(-1)^{a+b}c_2+(-1)^b c_3\right]. \nonumber
\end{equation}
and $|\beta_{ab}\rangle =\frac{1}{\sqrt{2}}[ |0,b\rangle +(-1)^a|1,1+b\rangle ] $ are the Bell states. The square root of the state $\rho^{BD}$ is 
\begin{align}
\sqrt{\rho^{BD}}=\frac{1}{4}\left[h \mathds{1}\otimes\mathds{1}+\sum^3_{i=1}d_i(\sigma^i \otimes \sigma^i)\right], \nonumber
\end{align}
where $h=\text{Tr}(\sqrt{\rho^{BD}})=\sum_{ab}\sqrt{\lambda_{ab}}$ and 
\begin{eqnarray}
d_1&=&\sqrt{\lambda_{00}}-\sqrt{\lambda_{01}}+\sqrt{\lambda_{10}}-\sqrt{\lambda_{11}}, \\ \nonumber
d_2&=&-\sqrt{\lambda_{00}}+\sqrt{\lambda_{01}}+\sqrt{\lambda_{10}}-\sqrt{\lambda_{11}}, \\ \nonumber
d_3&=&\sqrt{\lambda_{00}}+\sqrt{\lambda_{01}}-\sqrt{\lambda_{10}}-\sqrt{\lambda_{11}}. \\ \nonumber
\end{eqnarray}
Using Eq. (\ref{Aresult}), we compute the geometric discord for Bell diagonal state is 
\begin{align}
D(\rho^{BD})=1-\frac{1}{4}(h^2+\text{max}_j\{ d_j^2)\}.
\end{align}
\begin{figure*}[!ht]
\centering\includegraphics[width=0.8\linewidth]{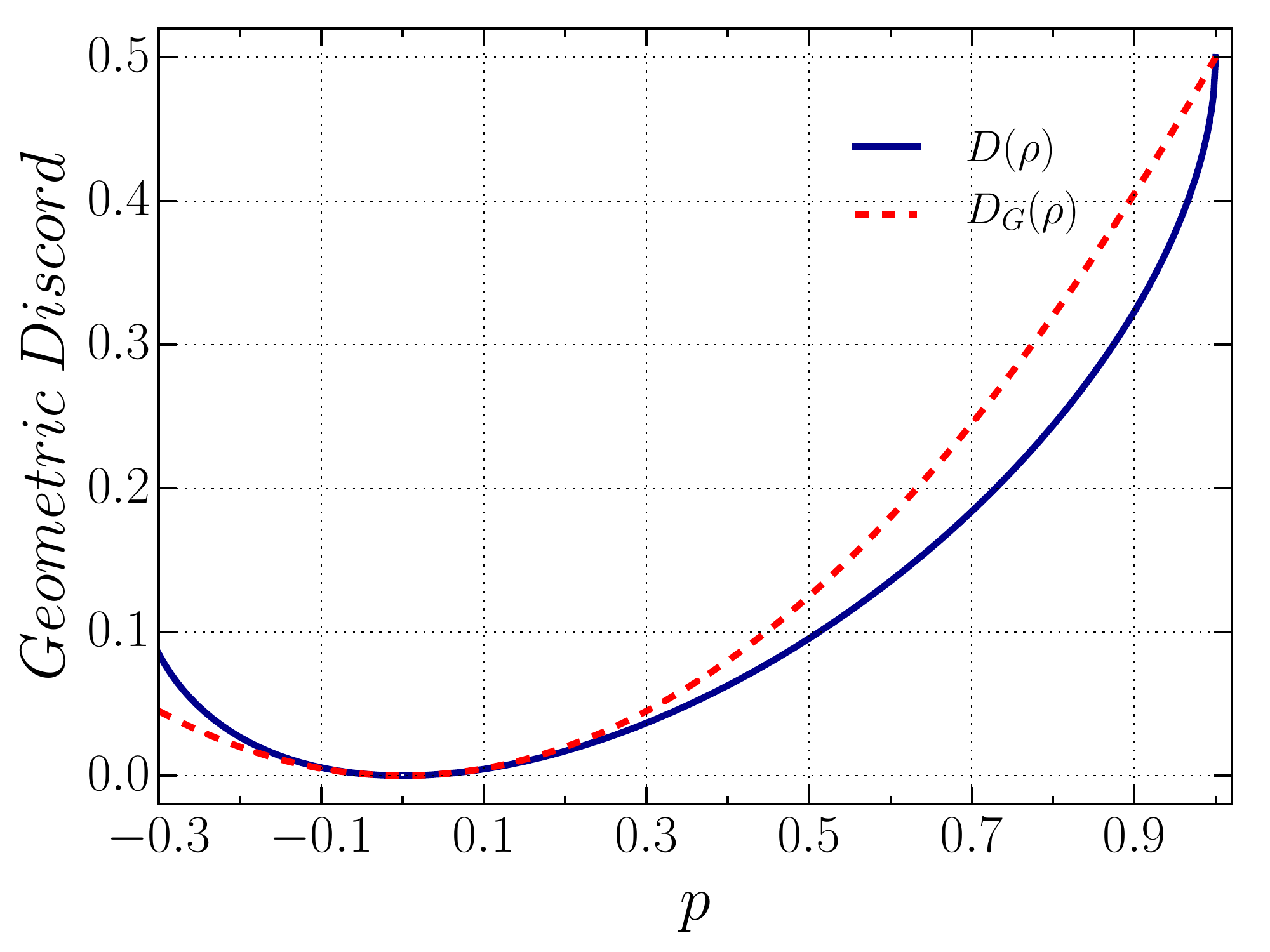}
\caption{(color online) Affinity and Hilbert-Schmidt norm based geometric discord for two qubit Werner state.}
\label{fig1}
\end{figure*}
In particular, if $c_1=c_2=c_3=-p$, then the Bell diagonal state reduced to two qubit Werner state,
\begin{align}
\rho^{BD}=\frac{1-p}{4}\mathds{1}+p|\Phi\rangle \langle \Phi|,  ~~~~~~~~~~ p\in[-1/3, 1], \nonumber
\end{align}
with $|\Phi\rangle \langle \Phi|$. The geometric discords of the Werner state are computed as 
\begin{align}
D(\rho^{BD})=\frac{1}{4}[ 1+p-\sqrt{(1-p)(1+3p)}],~~~~~~~~~~D_G(\rho^{BD}) =\frac{p^2}{2}.
\end{align}
To illustrate the relation between $D(\rho^{BD})$ and $D_G(\rho^{BD})$, we plot both the measures in Fig. (\ref{fig1}) and we observe that both the forms of geometric discord capture the quantumness of two qubit Werner state are qualitatively same. Further, both the quantities are vanishing at $p=0$ and maximum for $p=1$.

\textit{Werner state}: Next, we consider $m\times m-$ dimensional Werner state, which is defined as 
\begin{align}
\omega=\frac{m-\text{x}}{m^3-m}\mathds{1}+\frac{m\text{x}-1}{m^3-m}F,  ~~~~~~~~~~~~\text{x}\in[-1, 1],
\end{align}
with $F=\sum_{kl}|kl\rangle \langle kl|$. Affinity based geometric discord is computed as 
\begin{align}
D(\omega)=\frac{1}{2} \left(\frac{m-\text{x}}{m+1} -\sqrt{\frac{m-1}{m+1}(1-\text{x}^2)}\right) \nonumber
\end{align}
and the geometric discord \cite{Luo2010}
\begin{align}
D_G(\omega)=\frac{(m\text{x}-1)^2}{m(m-1)(m+1)^2}. \nonumber
\end{align}
It is observed that $D(\omega)=D_G(\omega)=0$, if and only if $\text{x}=1/m$. In the asymptotic limit $m\rightarrow \infty $,
\begin{align}
\lim_{m \to \infty}D(\omega)=\frac{1}{2}(1-\sqrt{1-x^2}),~~~~~~~~~~\lim_{m \to \infty} D_G(\omega) =0. \nonumber
\end{align}
The above equation suggests that the affinity-based geometric discord is more robust in higher dimension than  the Hilbert-Schmidt norm discord.

\textit{Isotropic state}: $m\times m-$ dimensional isotropic state is defined as 
\begin{align}
\rho=\frac{1-\text{x}}{m^2-1}\mathds{1}+\frac{m^2\text{x}}{m^2-1}|\Psi^+\rangle \langle \Psi^+|, ~~~~~~~\text{x}\in[0,1]
\end{align}
where $|\Psi^+\rangle=\frac{1}{\sqrt{m}}\sum_i |ii\rangle $. The affinity based discord is computed as 
\begin{align}
D(\rho)=\frac{1}{m} \left(\sqrt{(m-1)\text{x}}-\sqrt{\frac{1-\text{x}}{m+1}}\right)^2, \nonumber
\end{align}
and the another version is
\begin{align}
D_G(\rho)=\frac{(m^2\text{x}-1)}{m(m-1)(m+1)^2}. \nonumber
\end{align}
We see that $D(\rho)=D_G(\rho)=0$ if $\text{x}=1/m^2$. In the asymptotic limit,
\begin{align}
\lim_{m \to \infty}D(\rho)=\text{x},~~~~~~~~~~\lim_{m \to \infty} D_G(\rho) =\text{x}^2. \nonumber
\end{align}

\section{Conclusions}
\label{Concl}
In this article, we have proposed a new variant of geometric discord based on affinity metric and this quantity fixes local ancilla problem of Hilbert-Schmidt norm based quantity. We have computed analytically the closed formula of the proposed version of geometric discord for arbitrary pure state and $2 \times n-$ dimensional mixed state. Further, for any $m \times n-$ dimensional mixed state, the lower bound of affinity--based geometric discord is obtained in terms of eigenvalues of the correlation matrix. For illustration, the geometric discords  are computed for the well-known family of two qubit mixed states, namely Bell diagonal, isotropic and Werner states. 

Here, we mention that other correlation measure based on the Hilbert-Schmidt norm, such as measurement-induced nonlocality can also remedied in a similar way.
%
%

\begin{acknowledgements}

This work has been financially supported by the CSIR EMR Grant No. 03(1444)/18/EMR-II.
\end{acknowledgements}



\end{document}